High temperature $^1$H DOSY NMR reveals sourdough fermentation of wheat flour alters the molecular structure of water-extractable arabinoxylans.

Pasquinel Weckx[a,§], Víctor González Alonso[b,§], Ewoud Vaneeckhaute[a], Karel Duerinkcx[a], Luc De Vuyst[b,*], Eric Breynaert[a,c*]

[a.] NMR/X-ray platform for Convergence Research (NMRCoRe), KU Leuven, Celestijnenlaan 200F – box 2461, B-3001 Leuven, Belgium

[b.] Research Group of Industrial Microbiology and Food Biotechnology (IMDO), Faculty of Sciences and Bioengineering Sciences, Vrije Universiteit Brussel, Pleinlaan 2, B-1050 Brussels, Belgium

[c] Center for Surface Chemistry and Catalysis – Characterization and Application Team (COK-KAT), KU Leuven, B-3001 Leuven, Belgium

* Corresponding authors: eric.breynaert@kuleuven.be; luc.de.vuyst@vub.be

[§] P.W. and V.G.A. contributed equally to this work.

Abstract

Arabinoxylans are constituents of wheat flour that contribute to the dietary fiber properties of wheat. They exist in water-extractable and water-unextractable forms and contribute to human health. In bakery technology, especially the water-extractable arabinoxylans (WE-AX) are important due to their impact on viscosity and dough rheology. This study provides insights into the impact of wheat flour fermentation on WE-AX during sourdough production, offering potential applications for improving sourdough bread quality and its health benefits. The production of sourdoughs is known to increase the WE-AX fraction, yet the underlying (bio)chemical mechanisms remain unclear. This study investigated the alteration of WE-AX during the fermentation of wheat flour for sourdough production using $^1$H Diffusion Ordered SpectroscopY (DOSY) Nuclear Magnetic Resonance (NMR) at elevated temperature to analyze the structural changes of WE-AX during wheat flour fermentation for sourdough production with different lactic acid bacteria (LAB) strains. The results confirmed that DOSY NMR at elevated temperatures greatly improved the applicability of the method for analyzing larger biomolecules. Overall, a size reduction of the WE-AX compounds with increasing fermentation time was found. This was indicated both by the occurrence of higher self-diffusion coefficients, and increased transverse relaxation times. Further research is necessary to explain deviations from the general trend.

Introduction

Dietary fibre plays a significant role in human health. Its current intake in first world society is however low with respect to dietary guidelines,[1] especially in wheat bread-consuming communities. The viscous, water-extractable fibre fraction of wheat kernels can increase the viscosity of the digesta, thus modulating the glycaemic response and the re-absorption of bile acids.[2,3] Arabinoxylans (AX) are the main non-cellulosic dietary fibres in refined wheat flour, representing 60-70 % of the cell walls of the starchy endosperm.[4] Overall, AX are composed of a β-1,4-D-xylose backbone, substituted with α-L-arabinose moieties at the O-2 and/or O-3 positions.[4,5] Ferulic acid, in turn, may be linked to arabinose, allowing crosslinking between AX molecules.[6] Arabinogalactan peptide (AGP) is also present in wheat flour, influencing the analytical determination of AX.[7] One third of the AX fraction of flour can be considered as water-extractable and is classified as water-extractable arabinoxylans (WE-AX). Besides potential health benefits as dietary fiber, AX play a distinct role in bakery technology because of their impact on the water-holding capacity and binding of the dough, dough rheology, and starch retrogradation in bread.[4,8,9] During bread-making, the WE-AX increase the viscosity of the aqueous phase of the dough, thus stabilizing the liquid films around the gas cells. This, in turn, positively affects the bread quality by improving its foam texture.[10] It also has been shown that the production of wheat sourdoughs can solubilize a portion of the water-unextractable AX fraction, leading to an increased WE-AX fraction in both wheat[11–13] and rye doughs.[14] The molecular level impact of fermentation on the AX structure, especially in the flour-water mixtures used to produce sourdough, has however not been unravelled. It is also unknown whether all lactic acid bacteria (LAB) strains commonly encountered in sourdough production exert the same effect on the AX content and/or its molecular structure.

Liquid-state nuclear magnetic resonance (NMR) spectroscopy is a powerful method to analyze the molecular level structure of WE-AX, and has as of yet not been applied to monitor flour fermentation for sourdough production. NMR spectroscopy is inherently non-destructive, allowing multiple NMR experiments to be performed in succession, such that all desired information can be probed from a single sample. Diffusion Ordered SpectroscopY (DOSY) NMR virtually separates compounds in complex mixtures based on their self-diffusion, which is impacted by their structure, conformation, and molecular mass.[15] By implementing DOSY NMR, initial fractionation of the sample into different molecular mass fractions is no longer necessary.[16,17] In contrast with classical approaches, NMR spectroscopy not only allows the evaluation of the average molecular structure and molecular size distribution of the WE-AX; by the implementation of DOSY NMR, it also enables to assess if the fermentation induced structural changes (if any) are limited to specific molecular weight fractions or are occurring everywhere. Therefore, high-temperature $^1$H DOSY NMR should be investigated, in particular regarding flour fermentation for sourdough production.

As DOSY experiments are heavily and negatively impacted by convection, they are typically only conducted at room temperature to minimize temperature fluctuations and temperature gradients.[18–22] At room temperature (RT), the molecular mobility of dissolved WE-AX and thus the efficiency of averaging out dipolar interactions limits the resolution that can be achieved applying $^1$H NMR spectroscopy to WE-AX containing solutions at RT.[23]

In previous work, the potential of $^1$H DOSY NMR for analyzing WE-AX and AX oligosaccharides (AXOS) has been demonstrated.[16,17] A mixture of AXOS was separated virtually into distinct subpopulations, using room temperature DOSY and relaxation experiments to generate a three-dimensional space combining a diffusion, a $^{13}$C chemical shift, and a longitudinal relaxation dimension.[17] Experiments on WE-AX extracted from wheat flour and subsequently fractionated into three molecular mass fractions using an established ethanol precipitation protocol[24,25] have enabled the estimation of the diffusion coefficient of the smallest fractions of the WE-AX population using RT DOSY.[16] The low mobility of larger WE-AX fractions gives rise to increased residual dipolar interactions.[26] This in turn affects the transverse relaxation of nuclear spins and broadens the line-widths of the $^1$H and $^{13}$C NMR resonances, thereby preventing virtual separation of these fractions using RT DOSY NMR.[23]

The aim of the present study was to conduct the $^1$H DOSY NMR experiments at elevated temperature (323 K). The optimized, increased temperature will enhance the mobility of the WE-AX molecules, leading to improved (isotropic) averaging of dipolar couplings, thus inducing a narrowing of the NMR resonances. This improves the spectral resolution in the chemical shift dimension and allows the determination of diffusion coefficients for multiple subpopulations of the WE-AX pool. This approach will be applied to a WE-AX pool isolated from sourdough samples as a function of the fermentation time.

Materials and Methods

- <u>Wheat sourdough production</u>

Wheat lines derived from crosses of the Chinese spring wheat Yumai-34 (Y), a wheat variety with a high AX content,[27] with the European cultivar Ukrainka (U) were originally developed by the Centre for Agricultural Research (Agrártudományi Kutatóközpont, Martonvásár, Hungary). In the present study, YxU 069 and YxU 100 crosses, multiplied at Rothamsted Research (Harpenden, UK), were selected based on their total AX (TOT-AX) and WE-AX contents. They were grown at the TRANSfarm of the KU Leuven (Leuven, Belgium), and milled into white flour with an extraction rate of 67.0 %, a moisture content of 12.5 %, a TOT-AX content of 2.95 ± 0.35 % (on dry matter basis, dm), and a WE-AX content of 0.61 ± 0.04 % (dm). Xylanase activity was 1.85 xylanase units (XU)/g (dm) and was comparable to a reference commercial flour (Evina, 1.46 XU/g dm). Before use, the flour was stored at 7-10 °C in a humidity-controlled storage unit (N'ice holding cabinet, Irinox, Conegliano, Italy). The wheat kernels harvested were milled with an MLU-202 laboratory mill (Bühler, Uzwil, Switzerland) at the Laboratory of Food Chemistry and Biochemistry of the KU Leuven (Leuven, Belgium).

To produce wheat sourdoughs, wheat flour-water mixtures (50:50, m/m) were inoculated with four different LAB strains, thereby targeting 7.0 log of colony forming units (CFU) per g of sourdough as initial cell density, namely, *Limosilactobacillus fermentum* IMDO 130101 (LF, a laboratory wheat sourdough isolate),[28] *Lactococcus lactis* IMDO WA12L8 (LC, a laboratory high-AX wheat sourdough isolate; González-Alonso & De Vuyst, unpublished results), *Companilactobacillus crustorum* LMG 23699 (CC, an artisan bakery wheat sourdough isolate),[29] and *Companilactobacillus paralimentarius* IMDO BBRM18 (CP, a household rye sourdough isolate).[30] These fermentation processes were performed at 30 °C during 120 h. Sourdoughs were produced in duplicate; the duplicates were carried out two weeks apart from each other, with a freshly prepared inoculum. Samples were withdrawn after 0, 8, 48, and 120 h of fermentation, further referred to as time points t0, t8, t48, and t120, respectively. The sampling time points corresponded with the start (0 h), the middle of the exponential growth phase (8 h), the end of microbial growth (48 h), and the phase of microbial death (120 h) to find out if the microbial dynamics played a role regarding the WE-AX populations.

- <u>Isolation of water-extractable arabinoxylans</u>

The WE-AX were isolated from the sourdoughs of the first replicate, given the high reproducibility of both trials (data not shown). The isolation procedure was performed as described before,[16] with some modifications. After freeze-drying, 70 g of sourdough sample were weighed exactly and submitted to an ethanol reflux (80 %, v/v; Fisher Chemical, Hampton, New Hampshire, USA). Boiling stones (Acros/Thermo Fisher Scientific, Geel, Belgium) were added to prevent irregular boiling. After cooling, the ethanol was separated through rotary evaporation under vacuum at 35-40 °C. Water extraction was performed in duplicate with ultrapure water (MilliQ; Merck Millipore, Billerica, Massachusetts, USA), which was added (1:5, m/v) to the sourdough residues, followed by homogenization (150 rpm, 60 min, 4°C) using an orbital shaker RO 500 (Gerhardt, Königswinter, Germany). After centrifugation (3000 x $g$, 15 min, 20 °C), the supernatants were collected and incubated with α-amylase from *Bacillus licheniformis* (2,560 U; Sigma-Aldrich, Saint Louis, Missouri, USA) at 90°C for 30 min. Then, the mixtures were cooled to room temperature and centrifuged as mentioned above. The WE-AX were precipitated by adding ethanol gradually over 30 min under continuous stirring to obtain a final concentration of 65 % (v/v). The mixtures were then stirred for another 30 min and allowed to let the WE-AX precipitate overnight at 4 °C. The WE-AX were retrieved by centrifugation (10,000 x $g$, 30 min,

4 °C), dried in a laminar flow, and dissolved in MilliQ. Finally, the WE-AX mixtures were freeze-dried, and the WE-AX yields were determined gravimetrically and expressed as % of dry matter (dm).

- <u>Characterization of water-extractable arabinoxylans</u>

**Purity.** The purity of the WE-AX isolated was determined by gas chromatography with flame ionization detection, as described previously, allowing detection and quantification of arabinose, galactose, glucose, mannose, and xylose.[16] Briefly, 10-15 mg were weighed exactly and hydrolyzed with 5.0 ml of 2 M trifluoroacetic acid (Acros/Thermo Fisher Scientific), followed by reduction with a solution of borohydride (Sigma-Aldrich) under alkaline conditions, making use of ammonia (VWR International, Darmstadt, Germany) at 40°C for 30 min, and acetylation with acetic acid anhydride (VWR International). The AX concentrations and the arabinose over xylose (A/X) ratios were calculated as follows:

$$AX = 0.88 \times [(arabinose - 0.7 \times galactose)] + xylose) \quad (Eq.\ 1)$$
$$A/X = (arabinose - 0.7 \times galactose)/xylose \quad (Eq.\ 2)$$

**Molecular mass distribution.** The molecular mass (MM) distribution of the WE-AX isolated was determined with high-performance size exclusion chromatography (HPSEC) with refractive index detection, based on a method described previously.[16] A Shodex high-performance liquid chromatograph (Showa Denko KK, Tokyo, Japan) was used, equipped with a SIL-HT$_c$ autosampler (Shimadzu, Kyoto, Japan), Shodex SB-804 HQ (300 mm x 8 mm) and SB-806 (300 mm x 8 mm) columns connected in series, and a Shodex SB-G guard column (50 mm x 6 mm), in combination with a refractive index RID-10A detector operating at 40 °C. The column temperature was set at 30 °C. The mobile phase consisted of 0.3 % (m/v) NaCl in deionized water and its flow rate was set at 0.5 ml/min. Samples were prepared by dissolving 5.0 mg of WE-AX, taking into account the purity values as obtained above, in 1.0 ml of an aqueous solution of 0.3 % (m/v) NaCl. The samples were homogenized at 60 °C for 1 h, and then centrifuged (10,000 x *g*, 15 min, 20 °C). The supernatants were filtered (0.45 µm Millex filters; Merck Millipore) before injection into the columns (50 µl). Shodex P82 pullulan standards (MMs of $3.42 \times 10^2$, $1.32 \times 10^3$, $6.20 \times 10^3$, $1.00 \times 10^4$, $2.30 \times 10^4$, $4.88 \times 10^4$, $1.33 \times 10^5$, $2.00 \times 10^5$, $3.48 \times 10^5$, and $8.05 \times 10^5$ g/mol) were used. The peak areas obtained were expressed as arbitrary units.

**NMR analysis.** For the NMR analysis, aliquots of the freeze-dried WE-AX were accurately weighed, using a Sartorius CPA225D analytical balance (Sartorius, Göttingen, Germany) with a precision of 0.01 mg, and were dissolved in 1.0 mL of $D_2O$ (99.9 atom% D, Sigma-Aldrich) to obtain a WE-AX concentration of 1.0 mg/ml. Dissolution occurred in weighed containers, to allow efficient weight determination afterward. The samples were heated to 333 K for 60 min to promote dissolution. Following the dissolution step, the samples were centrifuged for 10 min at 2,733 x $g$ using a VWR Micro Star 12 (VWR International, Radnor, Pennsylvania, USA) centrifuge to promote rapid sedimentation of undissolved solid remains. The liquid supernatant (480 µL) was transferred to a non-spinning 5-mm NMR tube (Norell, Morganton, North Carolina, USA), in addition to 20 µL of a sodium trimethylsilylpropanesulfonate (DSS) solution of 25 mg/ml in $D_2O$. The remaining liquid was decanted, and the residual solid material was dried by exposure to $P_2O_5$ (Sigma-Aldrich). When dry, weighing the containers of each sample allowed for accurate mass determination of the residual solid material.

All $^1H$ NMR spectra were recorded with an 800 MHz Bruker spectrometer (Bruker Biospin, Billerica, Massachusetts, USA) at 323 K, operated by an AVANCE NEO console. The spectrometer, located in a temperature-controlled room (22 °C, maximum fluctuation of 0.1 °C over a 24 h time span) with a vibration-resistant floor, was equipped with a 5 mm multinuclear Bruker BBO probe ($^1H/^2H/X$) with active shielding and a maximum gradient strength of 4.45 G/mm. The $^1H$ NMR spectra were referenced at 0 ppm, using DSS (0.0173 ppm) present in the samples as a secondary reference.[31] The NMR experiments were all performed once, as the presence of the secondary reference allowed successive experiments to be checked for reproducibility. This also minimized the required experimental time, minimizing any effects of magnetic field drift or changes in field homogeneity, thus improving reproducibility and reliability of the NMR analysis.

1D $^1H$ NMR spectra were obtained with 64 scans, 12 ppm sweep width, 153,846 complex data points, a pi/2 excitation pulse at 16.4 kHz RF strength, and a relaxation delay of 15 s. Exponential line broadening (2 Hz), automatic phase correction, and baseline correction were performed using Bruker Topspin 4.1.4 software. Spectral decomposition was performed using DMfit software (CEMGTI, Orléans, France).[32] Lorentzian line shapes were used for the fitting.

The two-dimensional $^1H$ diffusion-ordered spectroscopy (DOSY) experiments were performed at 323 K. The DOSY plots were constructed by recording ten exponentially generated gradient amplitudes, varying from 2 to 95 % of the maximum gradient output (4.45 G/mm). Each slice was recorded with 16 scans, 12 ppm sweep width, 96,152 complex data points, a pi/2 excitation pulse at 16.4 kHz RF strength, and a relaxation delay of 15 s. The diffusion time and gradient pulse length were 125 and 2 ms, respectively, totaling 41 min of experimental time. The $^1H$ DOSY NMR data were processed with the General NMR Analysis Toolbox (GNAT, Manchester NMR Methodology Group, UK).[33] For extracting diffusion coefficients, a Gaussian line broadening of 3 Hz was used and monoexponential least-squares fitting with peak picking was performed.

Results and Discussion

**Purity of water-extractable arabinoxylans isolated from wheat sourdough productions as a function of the fermentation time**

The purity of the WE-AX isolated from samples taken during wheat sourdough productions was highest for the flour-water mixture at t0 (Table 1). The A/X ratio of these WE-AX decreased slightly upon fermentation. The lowest purities were obtained at t8, for which the glucose concentration was also highest (21.7-36.3 %). Free glucose, determined through acid hydrolysis of the WE-AX isolates, probably originated from larger glucose polymers that were inaccessible to prior amylase action, β-glucan co-extracted with AX being one of such possible polymers.[25,34]

**Table 1**. Water-extractable arabinoxylan (WE-AX) yield and purity, arabinose-to-xylose ratio (A/X), and glucose, galactose, and mannose concentrations of the WE-AX isolated from wheat sourdough productions initiated with four different lactic acid bacteria strains.

| Sourdough production | Time (h) | WE-AX yield (%) | WE-AX purity (%) | A/X ratio | Glucose (%) | Galactose (%) | Mannose (%) | HPSEC total peak area (arbitrary units x $10^6$) |
|---|---|---|---|---|---|---|---|---|
| FWM | 0 | 0.384 | 65.7 ± 1.2 | 0.53 ± 0.01 | 5.0 ± 0.3 | 3.5 ± 0.1 | 1.6 ± 0.0 | 5.8 ± 0.0 |
| LF | 8 | 0.681 | 40.0 ± 0.7 | 0.48 ± 0.00 | 35.2 ± 0.5 | 1.1 ± 0.0 | 0.3 ± 0.0 | 8.5 ± 0.1 |
| | 48 | 0.380 | 42.4 ± 0.5 | 0.42 ± 0.00 | 6.3 ± 0.1 | 2.5 ± 0.1 | 0.6 ± 0.2 | 6.1 ± 0.1 |
| | 120 | 0.540 | 50.6 ± 1.0 | 0.49 ± 0.00 | 5.5 ± 0.2 | 1.8 ± 0.1 | 0.7 ± 0.1 | 6.3 ± 0.1 |
| LC | 8 | 0.583 | 35.8 ± 0.4 | 0.51 ± 0.00 | 36.3 ± 0.5 | 1.7 ± 0.0 | 0.3 ± 0.0 | 9.3 ± 0.1 |
| | 48 | 0.547 | 55.9 ± 0.7 | 0.48 ± 0.00 | 5.9 ± 0.3 | 1.3 ± 0.0 | 0.3 ± 0.0 | 5.3 ± 0.0 |
| | 120 | 0.584 | 58.0 ± 1.0 | 0.47 ± 0.00 | 5.4 ± 0.4 | 1.6 ± 0.1 | 0.4 ± 0.0 | 5.5 ± 0.5 |
| CC | 8 | 0.489 | 45.7 ± 0.8 | 0.52 ± 0.00 | 21.7 ± 1.8 | 2.7 ± 0.1 | 0.4 ± 0.0 | 7.4 ± 0.2 |
| | 48 | 0.753 | 46.2 ± 1.1 | 0.47 ± 0.00 | 5.9 ± 0.7 | 2.0 ± 0.1 | 0.3 ± 0.1 | 5.8 ± 0.1 |
| | 120 | 0.831 | 46.5 ± 0.6 | 0.49 ± 0.00 | 4.6 ± 0.2 | 2.5 ± 0.1 | 0.3 ± 0.1 | 6.0 ± 0.1 |
| CP | 8 | 0.510 | 45.0 ± 1.0 | 0.52 ± 0.00 | 25.1 ± 1.1 | 2.8 ± 0.1 | 0.5 ± 0.0 | 7.5 ± 0.1 |
| | 48 | 0.654 | 51.1 ± 1.0 | 0.45 ± 0.00 | 5.8 ± 0.2 | 1.3 ± 0.1 | 0.4 ± 0.1 | 5.5 ± 0.1 |
| | 120 | 0.885 | 54.5 ± 0.6 | 0.45 ± 0.00 | 5.2 ± 0.1 | 1.0 ± 0.1 | 0.4 ± 0.0 | 5.2 ± 0.1 |

FWM, flour-water mixture; LF, *Limosilactobacillus fermentum* IMDO 130101; LC, *Lactococcus lactis* IMDO WA12L8; CC, *Companilactobacillus crustorum* LMG 23699; and CP, *Companilactobacillus paralimentarius* IMDO BBRM18.

**Molecular mass distribution of water-extractable arabinoxylans isolated from wheat sourdough productions as a function of the fermentation time**

The apparent molecular mass (MM) distribution of the WE-AX isolated as a function of the fermentation time, as determined by HPSEC, followed a similar profile for all wheat sourdough productions performed (Figure 1). Furthermore, the total peak area was consistently higher for the WE-AX isolated at t8, across all wheat sourdough productions performed. In general, the MM distribution decreased as the fermentation time increased. The highest intensity at 0 h was located between $400 \times 10^3$ and $800 \times 10^3$ g/mol, whereas it was closer to $400 \times 10^3$ g/mol after 8 h, at $200 \times 10^3$ g/mol after 48 h, and between $100 \times 10^3$ and $200 \times 10^3$ g/mol after 120 h of fermentation. The WE-AX profiles after 8 h of fermentation had a characteristic peak, corresponding with a MM ranging from 5 to $10 \times 10^3$ g/mol. In the case of a carbohydrate polymer made up of pentoses or hexoses, this would mean a degree of polymerization of 30-60 monomers. Given the high relative abundance of glucose in the WE-AX isolated after 8 h of fermentation, it was probable that these were glucose polymers rather than degradation products of AX. This peak was neither present at 48 h nor at 120 h of fermentation, but instead a smaller peak was then found, both in intensity and MM, corresponding with a size of 2-8 monomers. None of these small MM peaks were present at t0.

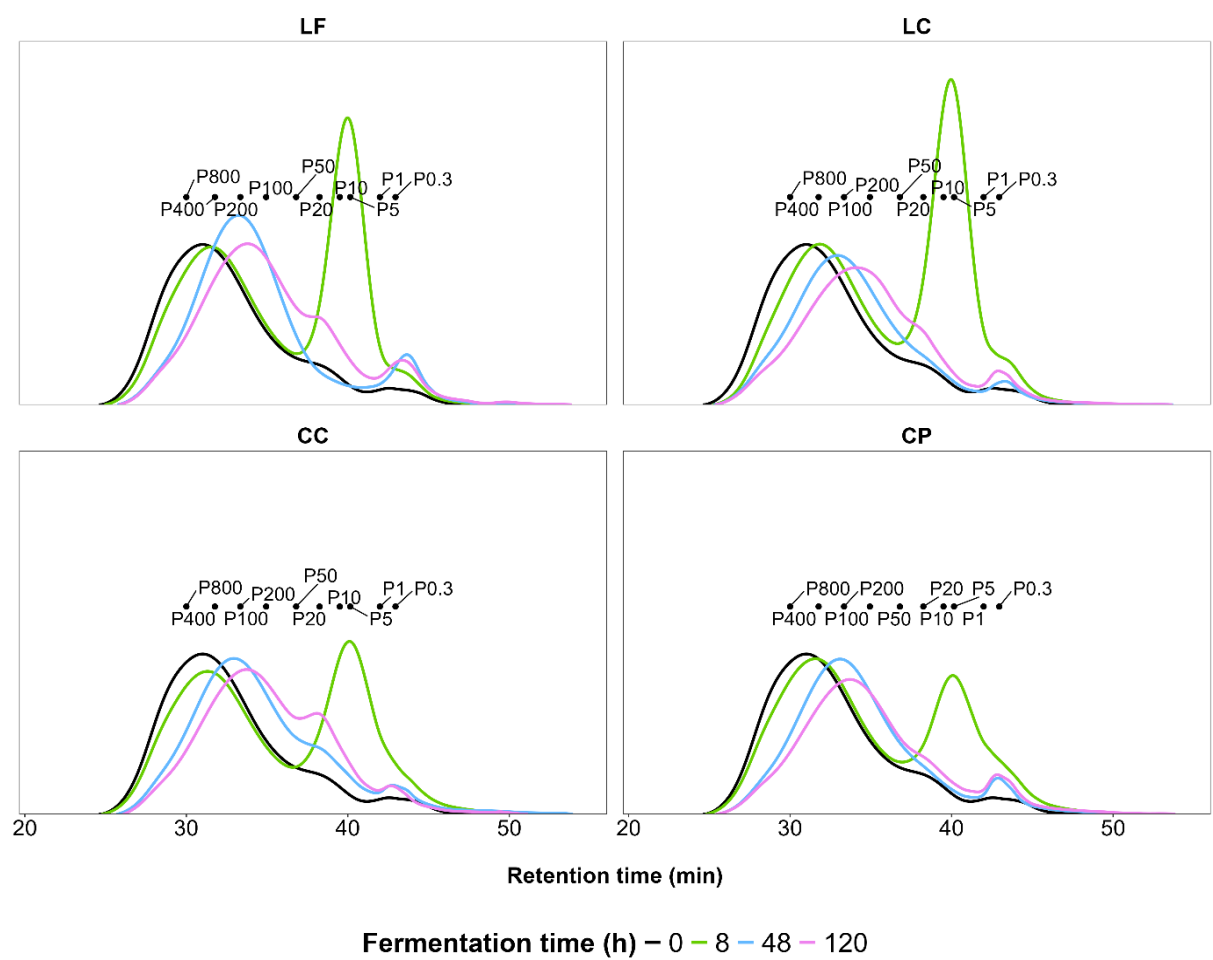

Fermentation time (h) — 0 — 8 — 48 — 120

**Figure 1.** Molecular mass distribution profiles (lines) of water-extractable arabinoxylans (WE-AX) isolated from wheat sourdough production samples taken as a function of the fermentation time (time points 0, 8, 48, and 120 h), and determined by high-performance size-exclusion chromatography (HPSEC) with refractive index detection. Pullulan standards of different molecular masses (dots), ranging from $8.05 \times 10^5$ to $3.42 \times 10^2$ g/mol (referred to as P followed by a number), are indicated as black dots. The wheat sourdough productions were initiated with four different lactic acid bacteria strains, namely, *Limosilactobacillus fermentum* IMDO 130101 (LF), *Lactococcus lactis* IMDO WA12L8 (LC), *Companilactobacillus crustorum* LMG 23699 (CC), and *Companilactobacillus paralimentarius* IMDO BBRM18 (CP).

### $^1$H DOSY NMR of water-extractable arabinoxylans

A recent publication employed 1-dimensional $^1$H NMR spectroscopy to detect changes in structure and composition of WE-AX in sourdough breads, comparing the state after milling of the flour and after baking of the dough.[35] $^1$H DOSY NMR spectroscopy until now has, however, only been performed on WE-AX samples extracted from flour samples.[16,36,37] This paper reports on NMR measurements of WE-AX directly isolated from wheat sourdough productions, taking samples as a function of the fermentation time. While $^1$H DOSY NMR measurements of WE-AX samples that have been reported in the literature were hitherto always carried out at room temperature, here the measurements were performed at elevated temperature (323 K). This enabled $^1$H DOSY experiments of the complete WE-AX population obtained from the wheat flour and wheat sourdough production samples, whereas previously this was only possible for the smaller subpopulations obtained after fractionation by ethanol precipitation.[16] The increased temperature increased molecular motion, increased tumbling of the WE-AX molecules, and decreased the viscosity of the solvent. Overall, this renders $^1$H DOSY experiments more manageable with standard hardware, especially for the larger WE-AX compounds. The enhanced (isotropic) averaging of dipolar couplings and its associated narrowing of the NMR resonances, increased the spectral resolution, and allowed more straightforward peak recognition and identification in the $^1$H DOSY NMR experiments.

This resolution increase is reflected in a comparison of the NMR spectrum of the t0 sample of the *Liml. fermentum* IMDO 130101-initiated wheat sourdough production at room temperature (295 K) and at the elevated temperature chosen for this study (323 K) (Figure 2). The most temperature-sensitive molecule present was HDO, the result of exchange between the trace amounts of $H_2O$ left in the WE-AX samples and the $D_2O$ solvent. The chemical shift of the WE-AX peaks was less sensitive to temperature. The biggest difference in the WE-AX spectrum was visible at 5.15 – 5.45 ppm, where the anomeric protons of the arabinose substitutions are located.[16] The elevated temperature resulted in an increased resolution and allowed a more evident analysis of this region. The increased distance to the HDO peak also facilitated this analysis, as the spectral baseline at the anomeric protons of arabinose portrayed a reduced curvature. There was a WE-AX peak located at 4.5 ppm, which overlapped with the H$_2$O peak at elevated temperature. However, this did not impact the subsequent analysis, since only the anomeric protons of the arabinose substitutions were analyzed.

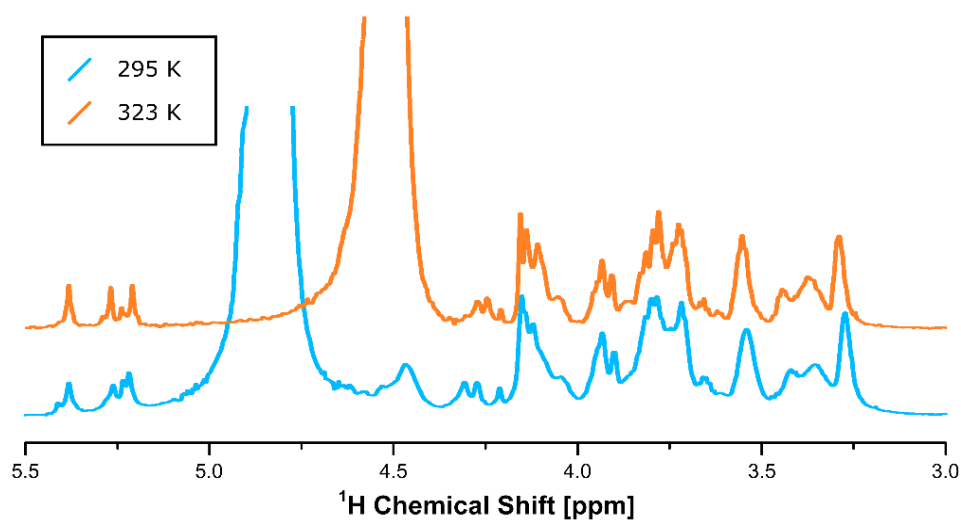

**Figure 2.** Visualization of the effect of temperature on the $^1$H DOSY NMR plots of the water-extractable arabinoxylans (WE-AX) isolated from wheat sourdough production samples, directly after inoculation of the lactic acid bacteria strain, *Limosilactobacillus fermentum* IMDO 130101 (LF), in the wheat flour-water mixture (t0). Blue data represent the measurement at 295 K; orange data represent the measurement at 323 K, the temperature chosen for this study.

The downside of DOSY NMR experiments at elevated temperature is that any temperature gradient present inside or close to the sample will induce convection, impacting the recorded resolution and diffusion coefficient.[18–22] To limit these effects, the NMR samples were free of solid particles and the temperature of the probe and the samples were allowed to stabilize for an extended period of time. NMR experiments at elevated temperature can also introduce instabilities in the shimming of the magnetic field, which in turn negatively affect the spectral resolution through broadening of the peaks. Preliminary tests in the system had indicated 323 K sufficiently high to induce the desired line narrowing, while it did not interfere with shim stability. During test measurements at 333 K, shim instabilities were observed, while the additional improvement of the linewidth was no longer significant.

## DOSY of WE-AX

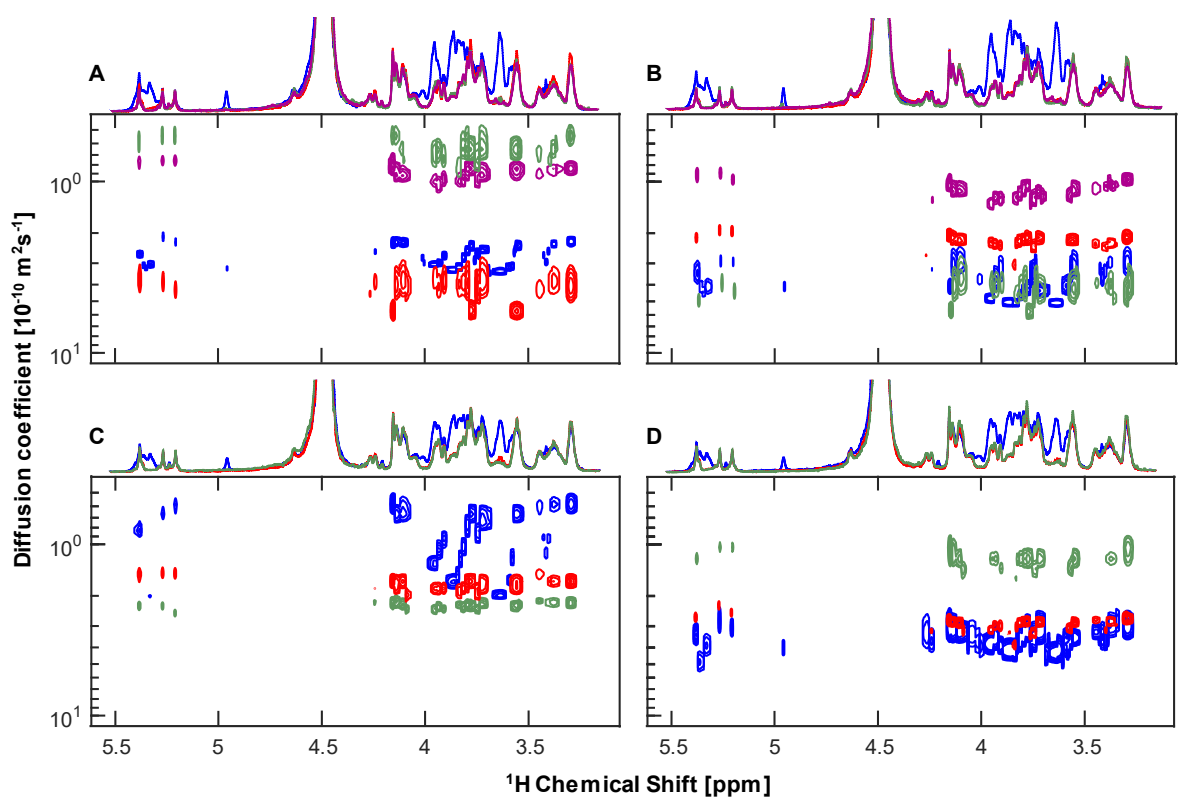

**Figure 3.** $^1$H DOSY NMR plots (recorded at 800 MHz and 323 K) of the water-extractable arabinoxylans (WE-AX) isolated from wheat sourdough production samples taken as a function of the fermentation time, namely, 0 h (purple), 8 h (blue), 48 h (red), and 120 h (green). The wheat sourdough productions were initiated with four different lactic acid bacteria (LAB) strains, namely, *Limosilactobacillus fermentum* IMDO 130101 (LF, A), *Lactococcus lactis* IMDO WA12L8 (LC, B), *Companilactobacillus crustorum* LMG 23699 (CC, C), and *Companilactobacillus paralimentarius* IMDO BBRM18 (CP, D). Each $^1$H DOSY plot is accompanied by the $^1$H 1D NMR spectra of the present samples.

Figure 3 shows the $^1$H DOSY NMR spectra of the WE-AX isolated from wheat sourdough production samples taken as a function of the fermentation time and initiated with four different LAB strains. The diffusion coefficient, set out on the vertical axis, contains larger values toward the bottom, indicating a faster diffusion. Figures 3A and 3B, showing the results of the *Liml. fermentum* IMDO 130101-initiated and *Lc. lactis* IMDO WA12L8-initiated wheat sourdough productions, include a t0 sample as well as the t8, t48, and t120 samples. Overall, the WE-AX isolated from the unfermented samples (0 h) were highly similar, both in diffusion coefficient as well as 1D NMR spectrum. As flour fermentation by different LAB strains could not have made an impact at t0, small differences between the t0 samples only resulted from experimental variability in the WE-AX extraction protocol and natural variations of the WE-AX structure. Since the differences were much smaller than the effects induced by the fermentation, there was no need to include additional t0 samples.

In the plot of the *Coml. crustorum* LMG 23699-initiated wheat sourdough production (Figure 3C), a general trend of increasing diffusion coefficient was visible as the fermentation time increased. This could be explained by a general decrease in size of the WE-AX compounds as fermentation proceeded. The *Lc. lactis* IMDO WA12L8-initiated wheat sourdough production (Figure 3B) agreed with this general trend, with the exception that the t8 sample showed a greater diffusion coefficient than the trend predicted. The diffusion coefficient of sample t8 also varied more over its chemical shift range compared to the other samples of the same wheat sourdough production process. The reason for this discrepancy was visible in the 1D NMR spectra, which showed that the t0, t48, and t120 samples overlapped greatly, whereas the t8 sample showed a different profile, with increased signal intensity mainly around 3.5–4.1 ppm and 4.9–5.5 ppm. These signals were not present in the t0 sample and were again absent in the t48 sample, indicating the emergence and disappearance of compounds during the fermentation process or isolation protocol.

The impact of these additional compounds on the $^1$H DOSY spectrum was that the peaks would behave as a summation of multiple compounds with different sizes, resulting in a diffusion coefficient indicating an estimated average for the multiple compounds present at that chemical shift. A possible explanation for the diverging behavior of the t8 sample in all LAB strain-initiated wheat sourdough productions was, therefore, that smaller compounds influenced the global diffusion coefficient of the WE-AX samples toward larger values. This effect took place for the four LAB strain-initiated wheat sourdough productions performed, indicating there might be an effect of the isolation procedure on the selection of compounds that were present in the t8 samples. Another explanation was that the fermentation at the 8-h fermentation time point indeed produced these small compounds, which all disappeared when 48 h of fermentation were passed. As discussed above, the t8 samples of all LAB strain-initiated wheat sourdough productions contained an additional amount of glucose structures, which explained the additional signals present in the NMR spectra (Figure 3) as well as in the HPSEC analysis (Figure 1).

The plots for the *Liml. fermentum* IMDO 130101-initiated (Figure 3A) and *Coml. paralimentarius* IMDO BBRM18-initiated (Figure 3D) wheat sourdough productions diverged, besides the t8 sample, in another way from the general trend of increased diffusion coefficient with increasing fermentation time. In both sets, the t120 sample showed a great decrease in diffusion coefficient compared with the other samples of their respective sets. The reason for this deviation was unclear, as it might be a result from the interaction of the isolation protocol with the compounds present after 120 h of fermentation. It could also depend on the different LAB strains and their interactions with the WE-AX compounds during fermentation, such as a possible production of an exopolysaccharide (EPS). The LAB strain *Liml. fermentum* IMDO 130101 is known for its production of EPS.[38] The presence of EPS would increase the viscosity of the sample, a functionality for which they are commonly employed in the food industry.[39,40] EPS has been reported to form supramolecular structures in the medium,[41] reducing the mobility of all the compounds in the sample, and thus leading to a decrease in the reported diffusion coefficient. However, the low concentrations at which this EPS would be produced do not allow its detection in the NMR spectra, considering their signals would overlap with the broad signals of the polysaccharides already present. Further research on the influence and presence of such possible EPS must be performed to confirm this hypothesis. One source that could be excluded was a possible instability of the $^1$H DOSY experiment, as was visualized in Figure 4.

**NMR DOSY of internal standards**

Figure 4 shows $^1$H DOSY NMR plots of DSS and HDO, measured simultaneously with the results shown in Figure 3. DSS was added as an internal standard for calibration of the spectrum, and HDO was present due to traces of $H_2O$ in the WE-AX samples, which exchanged with the $D_2O$ solvent. These signals served as control for the quality of the $^1$H DOSY NMR experiments, as their diffusion coefficient was not impacted by the different fermentation times or the isolation procedure of the WE-AX compounds. The stability of the DSS and HDO self-diffusivity observed across all samples for a single time point indicates that the high temperature $^1$H DOSY NMR experiment is a reliable tool, even for samples exhibiting resonances over a very broad chemical shift range and with largely different signal. In every WE-AX sample of all LAB strain-initiated wheat sourdough productions, the diffusion coefficient of DSS changed slightly with the WE-AX sample used. While the changes in self-diffusion coefficients observed for DSS were orders of magnitude smaller than those observed for the WE-AX components, the trend of the changes within the series of time-dependent samples for one sourdough production, mimicked the trend observed for the self-diffusion coefficients of the WE-AX compounds. This indicates small differences in the viscosity of the samples. The biggest difference, (±15%), was observed between the t120 and t48 samples of the *Lc. lactis* IMDO WA12L8-initiated wheat sourdough productions (Figure 4A). The same phenomenon was visible in the HDO signal, although the variations of the diffusion coefficient were even smaller, at most 5.35 % for the same samples mentioned above. In general, the addition of the internal standard indicated that the $^1$H DOSY NMR experiments at elevated temperature produced highly consistent data. No large discrepancies could be found in the signals belonging to DSS and HDO, ensuring that the $^1$H DOSY NMR plots of Figure 3 also gave a realistic representation of the WE-AX diffusion coefficients.

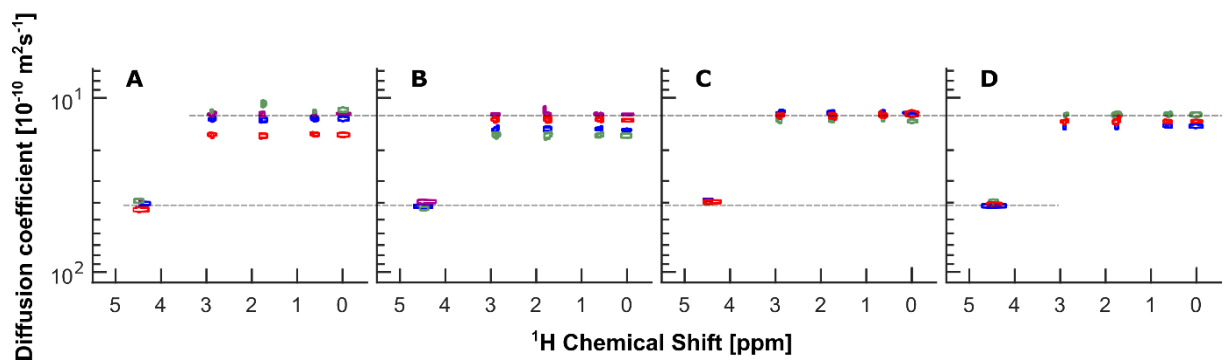

**Figure 4.** $^1$H DOSY NMR plots (recorded at 800 MHz and 323 K) of the sodium trimethylsilylpropanesulfonate (DSS, 0-2.9 ppm) and HDO (resulting from exchange between trace amounts of H$_2$O and the D$_2$O solvent, 4.5 ppm) signals recorded in the WE-AX isolated from wheat sourdough production samples taken as a function of the fermentation time, namely, 0 h (purple), 8 h (blue), 48 h (red), and 120 h (green). The wheat sourdough productions were initiated with four different lactic acid bacteria strains, namely, *Limosilactobacillus fermentum* IMDO 130101 (A), *Lactococcus lactis* IMDO WA12L8 (B), *Companilactobacillus crustorum* LMG 23699 (C), and *Companilactobacillus paralimentarius* IMDO BBRM18 (D).

**Explanation for the change in diffusion coefficient of the AX population**

A further analysis revealed the arabinose substitution patterns present in the WE-AX isolated, based on the NMR spectra of the anomeric protons of arabinose. Figure 5 shows the decomposition of the 5.15–5.45 ppm region of the 1D $^1$H NMR spectrum of the t0 sample of the *Liml. fermentum* IMDO 130101-initiated wheat sourdough production, in accordance with previous findings.[16] The decomposition revealed the presence of two major substitution patterns, being a mono-substitution of arabinose on the O-3 position of the xylose moiety in the backbone and a di-substitution on the O-3 and O-2 positions of the xylose moiety. These regions consisted of a large peak encompassed by a cluster of smaller peaks, representing slight differences in the neighboring structures of the polymer. A third, minor substitution pattern that occurred could be ascribed to the terminal α-linked arabinose in AGP.[42] The decomposition was done for all samples, except for the t8 samples, since the glucose present overlapped with the region of the arabinose anomers (seen in Figure 3), inhibiting the extraction of meaningful information about the substitution patterns.

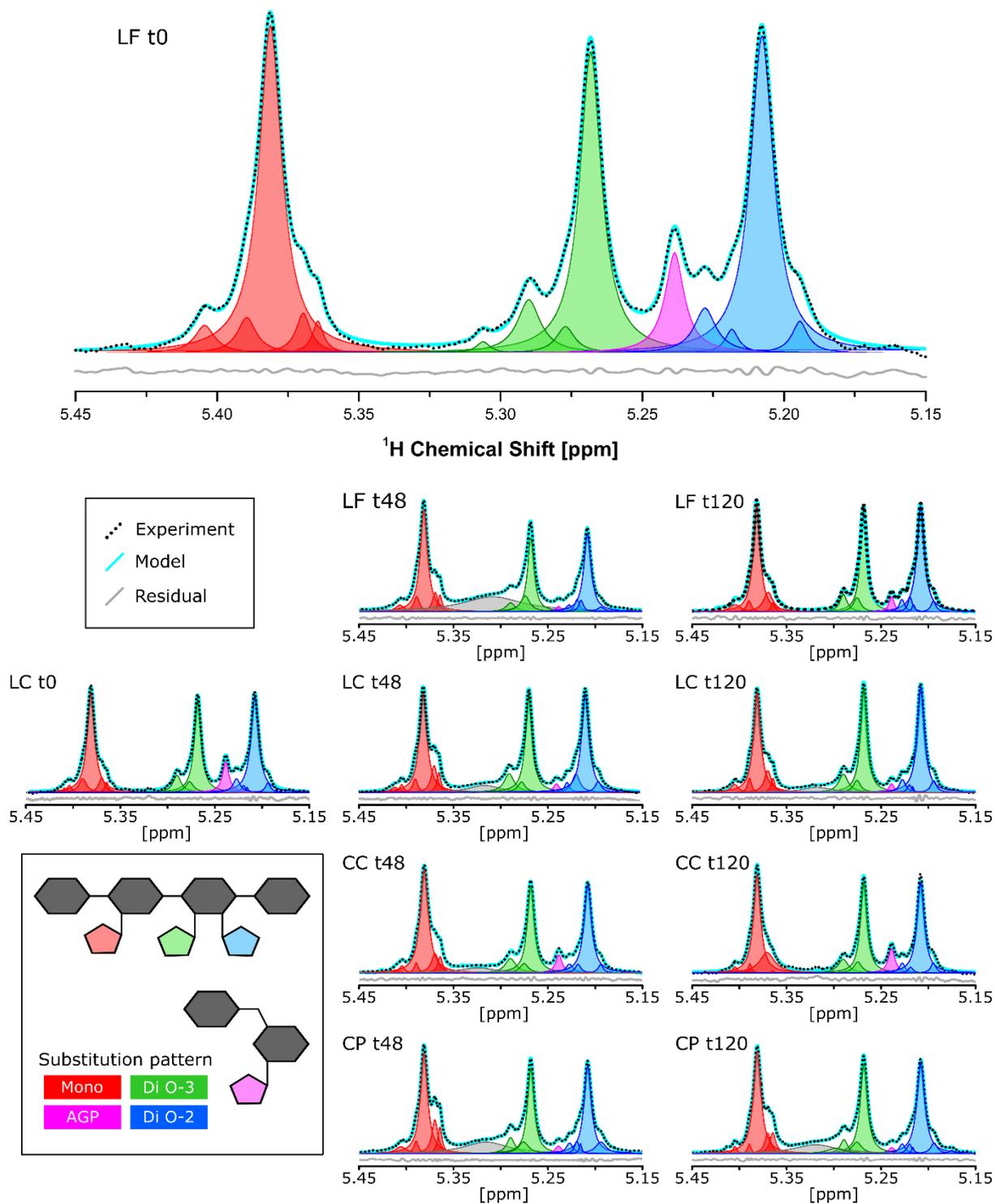

**Figure 5.** Decompositions of the 1-dimensional $^1$H NMR spectra (recorded at 800 MHz and 323 K) of the water-extractable arabinoxylans isolated from wheat sourdough productions, started with four different LAB strains, namely, *Limosilactobacillus fermentum* IMDO 130101 (LF, A), *Lactococcus lactis* IMDO WA12L8 (LC, B), *Companilactobacillus crustorum* LMG 23699 (CC, C), and *Companilactobacillus paralimentarius* IMDO BBRM18 (CP, D), as a function of the fermentation time (time points 0 h, 48 h, and 120 h). Only the region of 5.15-5.45 ppm is shown, as this region contains information about the arabinose substitutions on the xylose backbone chain. Peaks could be classified into three substitution patterns. The red region is accredited to a mono-substitution of arabinose on the O-3 position of xylose. The green and blue regions indicate a di-substitution of arabinose on the O-3 and O-2 positions, respectively. The purple region indicates a terminal α-linked arabinose in arabinogalactan peptide (AGP). Gray peaks are unknown compounds and are not included in the analyses.

The exact location and amount of the small peaks could vary between WE-AX samples, so the following analyses mainly discuss the large main peak for each substitution. The chemical shift of the main peaks did not change as a function of the fermentation time, nor with the LAB strain that was employed. This indicated a global uniform substitution pattern, independent of the fermentation time and LAB strain.

The widths of the NMR resonances give information about the transverse relaxation time constant of the magnetization ($T_2$).[23,43] The width of a resonance is affected by homogeneous and inhomogeneous broadening, which contribute to the overall transverse relaxation time constant $T_2^*$ through a homogeneous ($T_2$) and inhomogeneous ($T_2^+$) contribution, respectively (Eq. 3).[44]

$$\frac{1}{T_2^*} = \frac{1}{T_2} + \frac{1}{T_2^+} \text{ (Eq. 3)}$$

Narrow resonances imply a long $T_2$ relaxation time, originating from on average a smaller molecular structure in the mixture.

The NMR experiments were performed under strict conditions of temperature and shim stability, resulting in a uniform inhomogeneous broadening contribution for all samples. Sample-dependent changes in the value of $T_2^*$ of the resonance associated with a specific component, therefore, still provides information on the change in the underlying $T_2$ time constant.[23,43] $T_2^*$ relaxation time constants were derived from the linewidth of the Lorentzian fits (Eq. 4) of the main NMR resonances in the decomposed NMR spectra. In this equation, A represents the amplitude, $x_0$ the center frequency of the resonance, and $R^* = 1/T_2^*$ the overall transverse relaxation rate.

$$S(x) = \left(\frac{A\,R}{(x-x_0)^2 + R^{*2}}\right) \text{ (Eq. 4)}$$

The results are plotted in Figure 6. Overall, there was a slight decrease of the linewidths with the fermentation time, which, through association with an increase in the $T_2^*$ relaxation time constant, could indicate a decrease in average size of the WE-AX population. This observation coincided with the general trend observed in Figure 3, indicating smaller molecules and/or increasing molecular mobility with increasing fermentation time. The increase was, however, only significant in the *Liml.*

*fermentum* IMDO 130101 (Figure 6A) and *Lc. lactis* IMDO WA12L8-initiated wheat sourdough productions (Figure 6B). The narrowing of the resonances as fermentation time increased can only be explained by an increase in the $T_2$ relaxation time constant, resulting from a reduction in the size and/or mobility of the molecules.

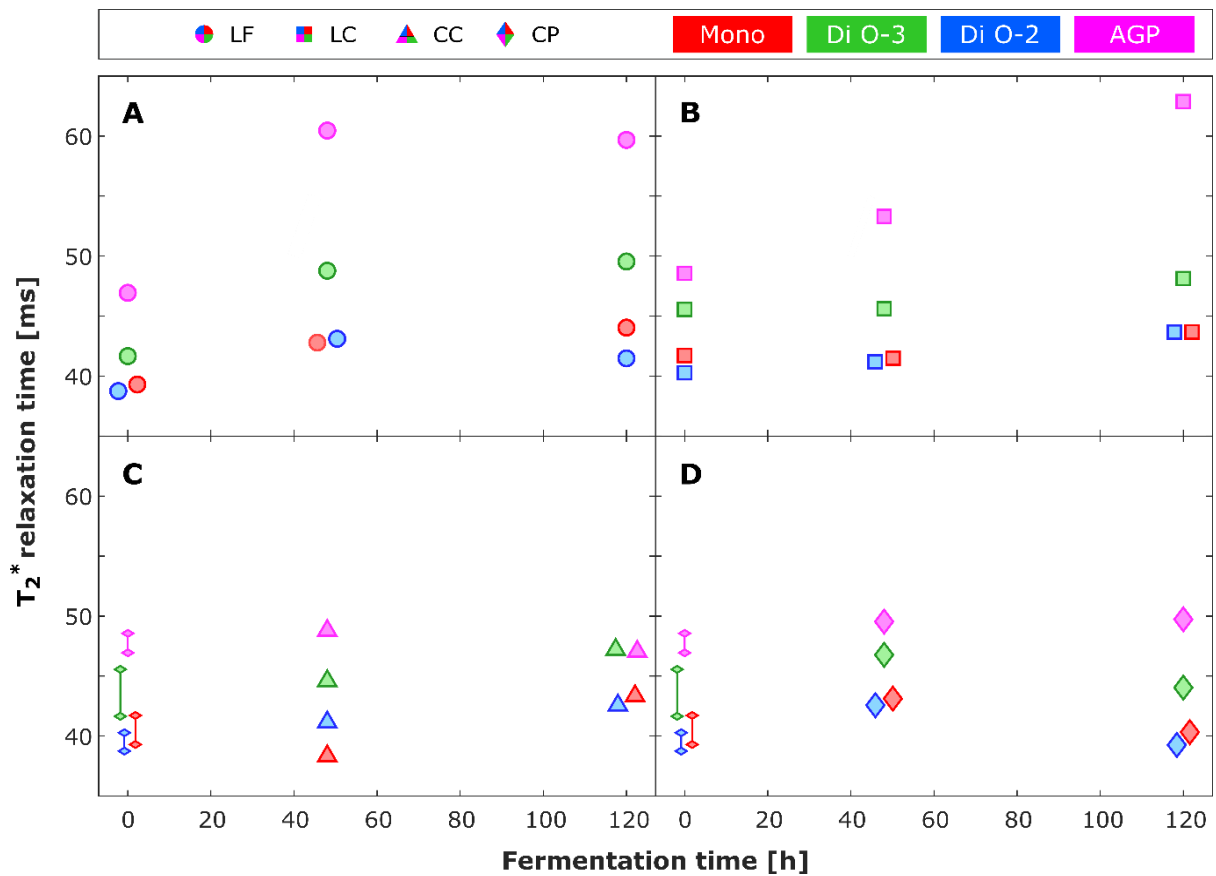

**Figure 6.** Estimate of the $T_2^*$ relaxation time constant for the main peaks of each substitution pattern, analyzed for each LAB strain-initiated wheat sourdough production employed at fermentation times t0, t48, and t120. The shape of the markers indicates the LAB strain employed, namely, *Limosilactobacillus fermentum* IMDO 130101 (LF, A), *Lactococcus lactis* IMDO WA12L8 (LC, B), *Companilactobacillus crustorum* LMG 23699 (CC, C), and *Companilactobacillus paralimentarius* IMDO BBRM18 (CP, D). The color indicates the substitution pattern. Overlapping data points are slightly separated horizontally for clarity, their vertical position remains unchanged. The colored lines in panes C and D represent the range of $T_2^*$ values measured for t0 in the samples shown in panes A and B. The 1-dimensional $^1$H NMR spectra underlying these fits were recorded at 800 MHz and 323 K.

A general note on the discrepancy between the observed trends for the analysis of the $T_2^*$ relaxation time constants and the diffusion coefficient analysis in Figure 3 is that the physical meaning of the $T_2^*$ relaxation time constant and the measured diffusion coefficient differ.[23] Self-diffusivity is directly correlated to the size of a molecule, implying larger molecules exhibit a lower self-diffusivity than smaller molecules. For transverse relaxation the simplified picture needs more nuance. Transverse relaxation is brought about by local fluctuations in the magnetic field 'felt' by a nuclear spin. These fluctuations result in random phase shifts and loss of transverse coherence. While $T_2^*$ typically increases with a decreasing molecular size, owing to higher mobility of the smaller molecules, in polymers such as WE-AX, local segment mobility can remain high for rather large molecules. In such case, $T_2^*$ can remain long while the self-diffusivity decreases.

It is peculiar to observe that the self-diffusivity coefficient and transverse relaxation time constant for the WE-AX compounds in the t120 sample of the *Liml. fermentum* IMDO 130101-initiated wheat sourdough production appear to be telling opposite stories. The DOSY measurements contrast the general trend of faster self-diffusion with increasing fermentation time, while the resonances still narrow, indicating smaller, more mobile components. Overall, this could mean that the WE-AX compounds could have aggregated into clusters with a high solvent content, such that the local segment mobility remained high, even though their apparent size increased, and associated with that their self-diffusivity decreased.

Finally, also the overall signal area on the resonances in the different spectral regions (mono, di O-2 and di O-3) were compared. Here, the full integration of all resonances in each spectral region was used, as the small peaks would have a considerable area to contribute to the substitution they represent. No general trend could be found in the relative area of the regions, indicating that the increase in the fermentation time did not induce significant changes in the overall substitution patterns. The AGP data points were the least reliable across fermentation time, since their intensity was the lowest of all substitution patterns, thus making it prone to large relative changes due to variations in the natural product and changes in the assignment and decomposition of the peaks. Important to note was that the O-2 and O-3 substitutions of the di-substituted pattern had a rough 1-to-1 ratio in all samples, indicating a realistic assignment of the small peaks in the decomposition.

**Analysis of the solid residue**

After dissolution of the WE-AX aliquots and subsequent centrifugation at 2,733 x *g*, the solid residue was weighed and expressed as a fraction of the total amount of initial sample. Figure 7A reveals a general increase in solid residue with increasing fermentation time. This coincided with the fact that the WE-AX isolated from fermented samples were of a lower purity than those obtained from unfermented samples (Table 1).

By plotting the data in a different way, Figure 7B revealed additional information. Here, the extracted liquid mass was compared to the purity of the samples. The diagonal showed the theoretical maximum of extraction if the samples would consist only of soluble WE-AX compounds and insoluble impurities. In general, the data points were placed close to this diagonal, indicating a high extraction effectiveness and, hence, a solid residual highly depleted of soluble WE-AX compounds. One exception was the cluster of blue data points of the t8 samples, which exceeded the diagonal drastically. This exception was expected due to the presence of glucose-based polymers, already noticed during the purity analysis as well as in the 1D NMR spectra.

A possible explanation for the increase of solid residual with increasing fermentation time could be the aggregation of larger WE-AX compounds as fermentation proceeded. This would lead to a reduced average size of the soluble WE-AX compounds, while not affecting the purity reported in Table 1. This hypothesis would also explain the trends observed in the diffusion coefficients in Figure 3 and the $T_2^*$ relaxation time constant analysis of Figure 6.

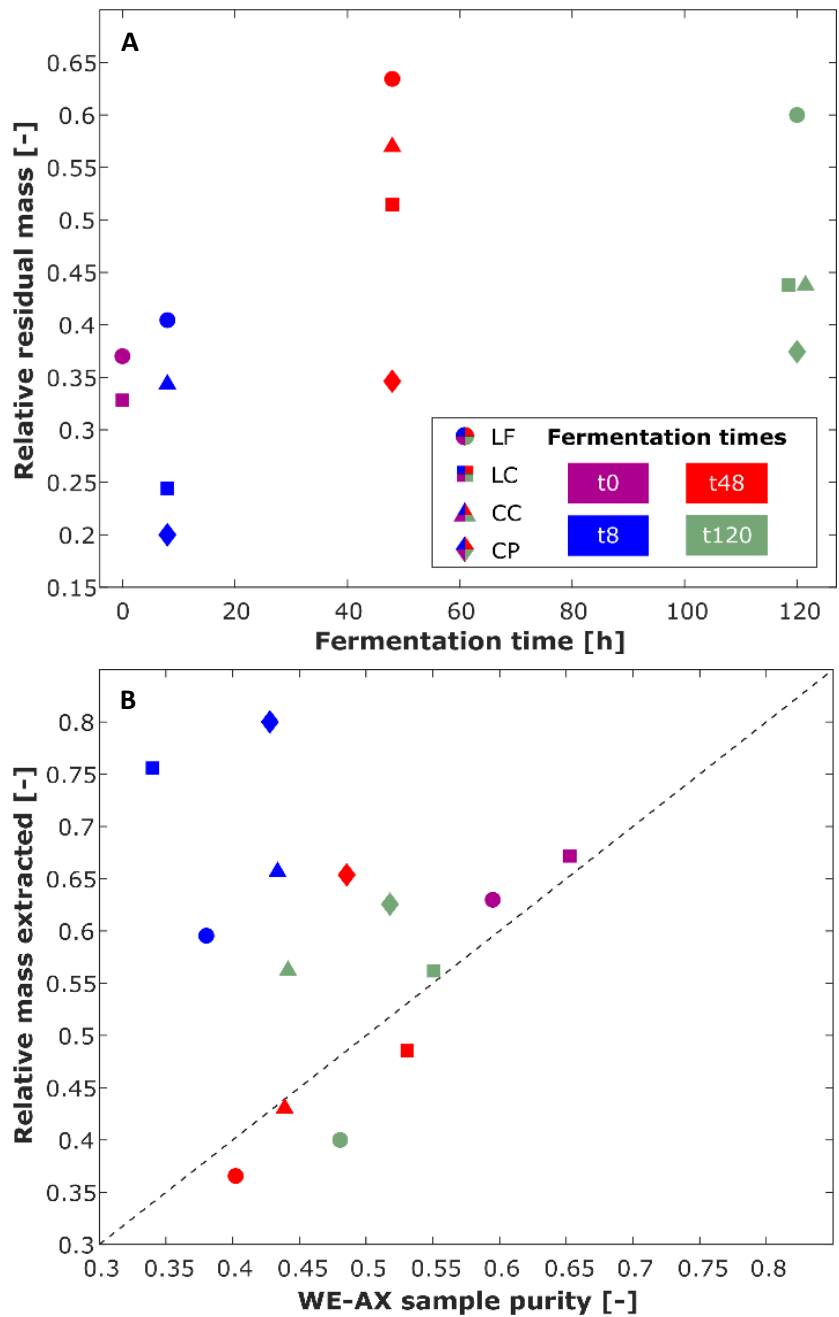

**Figure 7.** Data obtained from the solid residuals of the WE-AX samples. The shape of the markers indicates the lactic acid bacteria strains employed, namely, *Limosilactobacillus fermentum* IMDO 130101 (LF), *Lactococcus lactis* IMDO WA12L8 (LC), *Companilactobacillus crustorum* LMG 23699 (CC), and *Companilactobacillus paralimentarius* IMDO BBRM18 (CP). The color indicates the fermentation time (time points 0 h, 8 h, 48 h, and 120 h).

**7A:** Overview of the relative residual masses left after extraction and drying. Overlapping data points were slightly separated horizontally for clarity.

**7B:** Overview of the mass percentage of the samples extracted by dissolution, compared to the initial WE-AX purity of the samples. The dashed diagonal shows a theoretical maximum extraction limit if any impurities would be insoluble.

**Conclusion**

This research showed that NMR spectroscopy at elevated temperature allowed for an estimation of the diffusion coefficients of complete WE-AX populations by enhancing their mobility. This procedure was applied to WE-AX samples isolated from four wheat sourdough productions, inoculated with different LAB strains as starter culture. Sampling at different time points during the fermentation process allowed for an analysis of how the fermentation impacted the diffusion coefficient of the WE-AX compounds as well as the individual substitution patterns.

$^1$H DOSY NMR experiments revealed that there was a general trend of increasing diffusion coefficient, and thus reduction of the molecular size, as fermentation time proceeded. Deviations from this trend might be explained through the possible presence of sugar polymers, for instance an EPS, which might form clusters at high fermentation times, artificially lowering the observed diffusion coefficient. However, more experiments on the influence and presence of such an EPS must be performed to confirm this hypothesis.

Decomposition of the 1D $^1$H NMR spectra revealed insights into the influence of the fermentation time on the three arabinose substitution patterns of the WE-AX compounds. No trends were identified in the effect of the fermentation time on the chemical shifts or areas of the peaks representing the different arabinose substitution patterns. There was, however, a trend of increasing $T_2^*$ relaxation time as fermentation time proceeded, indicating a slight decrease of the overall size of the WE-AX molecules. This finding corresponded with the earlier trend observed in the diffusion coefficient data. Further, the analysis of the solid residuals of the WE-AX samples showed an increase as fermentation time proceeded. This, combined with the overall trend of size reduction of the WE-AX compounds, might indicate that large WE-AX compounds became insoluble as fermentation proceeded. More experiments must be done to confirm this hypothesis.

Finally, the data reported above provided insights into the impact of wheat flour fermentation on WE-AX during sourdough production. This could offer potential applications for improving sourdough bread quality and its health benefits when fully understanding the variation of the WE-AX size and structure.

**Acknowledgements**

This work was financially supported by the Research Council of the Vrije Universiteit Brussel (SRP7 and IOF3017 projects) and a Strategic Basic Research project (cSBO-Fibraxfun, HBC.2018.0505) from the Flemish Agency for Innovation and Entrepreneurship (VLAIO) and the spearhead cluster Flanders' FOOD. VGA was the recipient of a PhD fellowship from the Vrije Universiteit Brussel (cSBO Fibraxfun). NMRCoRe is supported by the Hercules Foundation (AKUL/13/21), by the Flemish Government as an international research infrastructure (I001321N), and by the Department EWI via the Hermes Fund (AH.2016.134). The authors would like to thank Wannes De Man for his support with the WE-AX isolations and J. Delcour for coordinating the Fibraxfun cSBO project.